\documentstyle[12pt]{article}
\def\beqar {\begin{eqnarray}}
\def\eeqar {\end{eqnarray}}
\def\beq {\begin{equation}}
\def\eeq {\end{equation}}
\def \ep {{\epsilon}}

\def \l {{\lambda}}
\def \la {{\langle}}
\def \ra {{\rangle}}

\def \Tr {{\rm Tr}}
\def \bz {{\bar z}}
\def \S {{\cal S}}
\begin{document}

\begin{titlepage}
\null\vspace{-62pt}

\pagestyle{empty}
\begin{center}
\rightline{}
\rightline{CCNY-HEP-01-01 }
\rightline{RU-01-3-B}
\rightline{hep-th/0102181}

\vspace{1.0truein} 
{\large\bf On Level Quantization for the Noncommutative }\\
\vskip .05in
{\large\bf Chern-Simons Theory}\\

\vspace{1in} V.P. NAIR$^a$ and A.P. POLYCHRONAKOS$^{b,}$
\footnote{On leave from Theoretical Physics Dept., Uppsala
University, Sweden and Physics Dept., University of Ioannina,
Greece.}\\
\vskip .1in {\it $^{a,b}$Physics Department, City College of the
CUNY\\ New York, NY 10031\\
\vskip .05in
$^{a,b}$The Graduate School and University Center\\ City
University of New York\\ New York, NY 10016\\
\vskip .05in
$^b$Physics Department, Rockefeller University\\ New York, NY
10021\\}
\vskip .05in {\rm E-mail: vpn@ajanta.sci.ccny.cuny.edu, 
poly@teorfys.uu.se}\\
\vspace{1.in}
\centerline{\bf Abstract}

\end{center}
\baselineskip=18pt

We show that the coefficient of the three-dimensional
Chern-Simons action on the noncommutative plane must be
quantized. Similar considerations apply in other dimensions as
well.

\end{titlepage}

\hoffset=0in
\newpage
\pagestyle{plain}
\setcounter{page}{2}
\newpage

Chern-Simons field theories have been
extensively investigated in various contexts since their
appearance in physics literature as topological mass terms for
odd dimensional gauge theories \cite{DJT}. Chern-Simons theories on
noncommutative spaces were introduced recently using the 
star-product \cite{star} or the operator formulation \cite{poly}
and there have
been a number of papers investigating the properties of such
theories \cite{general}. In the commuting case, it is well known that
invariance of the theory under gauge transformations which are
homotopically nontrivial requires the quantization of the
coefficient of the CS term in the action, the so-called level
number. An immediate and natural question is whether such a
quantization would hold on a noncommuting space as well; this is
the subject of this paper. This question was recently addressed
in reference \cite{SJ}, where it was argued that there is no quantization
of the level number for the noncommutative (NC) plane. We show that
there is actually quantization of the level number on the plane.
In fact the result is stronger in the noncommutative case: there
is quantization even for the $U(1)$-theory. Consistency with the
commutative limit is obtained in the following way.
As the noncommutativity parameter
$\theta$ approaches zero, the relevant transformations go over to the
smooth homotopically nontrivial transformations in $SU(N)$. However, 
if we interpret them as $U(1)$-transformations, the limit is singular. 
There is then no need to demand invariance under these for the
$U(1)$-theory in the commutative limit, removing the reason for
level quantization in this limit.

We follow the notation of reference \cite{poly}. The CS action in three
dimensions is given in terms of covariant derivative operators
$D_\mu$ by
\beq
\S = \l 2\pi \theta \int dt ~\Tr ~\left( i{2\over 3} D_\mu D_\nu
D_\alpha  +\omega_{\mu\nu} D_\alpha \right)\ep^{\mu\nu\alpha}
\label{1}
\eeq
Here space consists of a noncommutative plane $x^1 , x^2$ and a 
commutative third dimension $x^0 = t$, satisfying
\beq
[x^\mu ,x^\nu ]= i\theta^{\mu\nu}
\eeq
with the antisymmetric $\theta^{\mu\nu}$ and $\omega_{\mu\nu}$ 
defined in terms of a c-number parameter $\theta$ as
\beqar
\theta^{12} &&=-\theta^{21} =\theta ~,~~~ 
\theta^{01}=\theta^{02}=0 \nonumber \\
\omega_{12} &&=-\omega_{21} = -\frac{1}{\theta} ~,~~~
\omega_{01}=\omega_{02}=0
\eeqar
A standard realization of the NC
plane is given by the Fock oscillator basis 
$\vert n \ra$, $n=0,1,\cdots$, on which $D_1$ and $D_2$ act as
arbitrary hermitian $t$-dependent operators, while
\beq
D_0 = -i \partial_t + A_0
\eeq
with $A_0$ a hermitian $t$-dependent operator. $U(1)$ and $U(N)$
gauge theory are recovered as different embeddings of the noncommutative
coordinates in the oscillator space. Specifically, taking the direct
sum of $N$ copies of the Fock space (which is isomorphic to
a single space via $\vert Nn+a \ra \sim \vert n,a \ra$) 
we can realize the coordinates $x^1 , x^2$ as ladder operators, i.e.,
\beq
( x^1 + i x^2 ) \vert n,a \ra = \sqrt{2\theta n} ~\vert n-1 ,a \ra
~,~~~~~~ a=1, \dots N
\eeq
The noncommutative partial derivative operators become
\beq
\partial_j = i \omega_{jk} x^k ~,~~~~~~j,k=1,2
\eeq
which indeed generate translations of $x^j$ upon commutation.
$\Tr$ in (\ref{1}) is understood as the trace in the full 
oscillator space; $2\pi \theta$ times trace over $n$ represents
integration over the noncommutative plane while the remaining
trace over $a$ represents $U(N)$ group trace.

Using the explicit form of the operator $D_0 = -i\partial_t + A_0$
the last term of (\ref{1}) is $\Tr (\omega A_0)$ and the action
becomes
\beq
\S =  \l 2\pi \theta \int dt~ \Tr ~\left( i{2\over 3} D_\mu D_\nu
D_\alpha \right)\ep^{\mu\nu\alpha} ~+~ 4\pi \l \int dt~ \Tr
(A_0) \label{2}
\eeq 
Notice that the last term has the form of a one-dimensional
CS action. 

Gauge transformations act on the fields as
\beq 
D_\mu \rightarrow U ~D_\mu ~U^{-1}\label{3}
\eeq
where $U$ is a $t$-dependent unitary transformation with the 
property that $U$ acts as identity on the states
$\vert n\ra$ of the oscillator basis as $n \rightarrow \infty$.
This property is the noncommutative version of the requirement
that gauge transformations go to the identity at spatial
infinity. For the level quantization argument, we consider
$U$'s which also become trivial at time infinity, that is,
$U\rightarrow 1$ as $t\rightarrow \pm \infty$. In the
commutative case, these requirements tell us that the maps
$U:{\bf R}^3 \rightarrow G$ are equivalent to the maps
$U : S^3 \rightarrow G$ and are classified by the winding number
of the homotopy group $\Pi_3(G)$. 

Under the unitary transformation (\ref{3}), the change in the
action (\ref{2}) is given by
\beq
\Delta \S = i 4\pi \l \int dt~ \Tr ( {\dot U} U^{-1} ) \label{4}
\eeq 
Notice that with $U$ acting as identity on states $\vert n\ra$ 
for large $n$,
$D_\mu$ do not change for large $n$, the cyclic symmetry of the trace 
holds and the first term in (\ref{2}) remains invariant.
To see that the change of action in (\ref{4})
can indeed produce a nontrivial result,
consider first $U$'s of the form
\beq
U\vert n\ra = \vert n \ra ~~~{\rm for}~~~n \geq N.
\eeq
Then $U$ is essentially
a $U(N)$-matrix  and the integral $\int dt~\Tr ({\dot U}
U^{-1})$ is the winding number for 
$\Pi_1 (U(N))$. Specifically, we can write $U= e^{i\alpha (t)}
~V$ with $\det V =1$, i.e.,
$V$ is an element of $SU(N)$. Then ${\dot V} V^{-1}$ is traceless
and 
\beq
\Tr ({\dot U} U^{-1}) = i N {\dot \alpha} (t)
\eeq
Since $\exp(-2\pi i /N) {\bf 1}$ is a central element of $SU(N)$, 
and $\Pi_1 (SU(N))$ is trivial, the periodicity of $\alpha$ is 
$2\pi /N$. With $U\rightarrow 1$ as $t\rightarrow \pm \infty$, 
the change in $\alpha$ from $t=-\infty$ to $t=+\infty$, namely 
$\Delta \alpha = \alpha (\infty ) -\alpha (-\infty )$, must be
an integral multiple of 
$2\pi /N$. The change in the action (\ref{4}) is given by
\beqar
\Delta \S &&= 4\pi \l N \int dt ~{\dot \alpha} \nonumber\\ &&=
4\pi \l N \Delta \alpha = 8\pi^2 \l m \label{5}
\eeqar 
where $m$ is an integer. Setting this to be an integral
multiple of $2\pi$ for single-valuedness of $\exp(i\S )$, we
find 
\beq 
4\pi \l =k
\eeq
where $k$ is an integer. The coefficient of
the CS action is thus quantized. The quantization is independent
of $N$; the specific value of $N$ is immaterial for the
argument. The quantization is also independent of $\theta$ and
conforms to the quantization of the commutative non-Abelian
Chern-Simons coefficient.

The above argument applies to the $U(1)$-theory as well as 
the non-Abelian theory. The difference between these theories in 
the commutative limit arises from the different behavior of
the limit of $U$ as $\theta$ goes to
zero, as we will argue shortly. The argument is also, in essence, 
the argument for the
level number quantization of a one-dimensional commutative CS
action. The noncommutative CS action in higher dimensions also
contain a term proportional to the one-dimensional CS action and
hence a similar quantization of the level number holds in
higher dimensions as well. Specifically, the action in $2n+1$
dimensions is \cite{poly}
\beq
\S = \l \sqrt{\det(2\pi \theta)} ~\Tr \sum_{k=0}^n {n+1 \choose k+1}
\frac{k+1}{2k+1} \omega^{n-k} {\rm D}^{2k+1}
\label{n} \eeq
In the above, $\theta$ and $\omega$ are again the antisymmetric
two-tensor and its inverse two-form specifying the noncommutativity
of space, while ${\rm D} = D_\mu dx^\mu$ is the operator one-form
of covariant derivatives. The $k\neq n$-terms in (\ref{n}) correspond to 
lower-dimensional Chern-Simons forms; their coefficients are
chosen to reproduce the standard result upon substituting
$D_\mu = -i\partial_\mu + A_\mu$. The presence of the one-dimensional
term $\omega^n D \sim \sqrt{det \omega}~ A_0$ is particularly crucial:
in its absence, the equations of motion arising from (\ref{n}) would
admit $D_\mu = 0$ as a solution and would not reproduce ordinary
noncommutative space. An argument similar to the one presented
for the three-dimensional case demonstrates that all higher terms in
(\ref{n}) remain invariant under gauge transformations that approach
the identity at infinity, while the one-dimensional term acquires
a contribution proportional to the winding number of an effective
$U(N)$ transformation. A quantization of the level $\l$ follows,
in accordance with the commutative nonabelian result.

It is instructive to demonstrate the arguments given above with
an explicit example in the three dimensional case.
A concrete nontrivial unitary transformation $U$ can be given as
follows. Consider two copies of the oscillator Fock basis $\vert
n,a\ra$, $ a=0,1$. We write $U$ in the $2\times 2$-form
\beq 
U_{ab}= \left(\matrix{ \sum_n A_n \vert n,1\ra \la n,1\vert
&\sum_n B_n \vert n,1\ra
\la n-1, 2\vert \cr {}&{}\cr
\sum_n B_n \vert n-1, 2\ra \la n,1 \vert & \sum_n A^*_{n+1}
\vert n,2\ra \la n,2\vert \cr}
\right)\label{6}
\eeq 
where
\beqar 
A_n &&= {2\theta n - (\rho +it )^2 \over 2\theta n +\rho^2
+t^2 }\nonumber\\ 
B_n &&= -~{2i\rho \sqrt{2\theta n} \over 2\theta n
+\rho^2 +t^2 }\label{7}
\eeqar 
The two copies of the Fock basis may be considered as a
single space with the identification
\beq
\vert n,a \ra = \vert 2n +a \ra \label{8}
\eeq 
In other words, we are simply splitting the oscillator Fock
space into the even and odd subspaces to write $U$ in the
$2\times 2$-form.  Notice that for $n \gg \rho^2 /\theta$,
$U\approx {\bf 1}$, as required. $\rho$ is the scale size of how
much this transformation differs from the identity.

We can once again calculate the integral $\int dt ~\Tr ({\dot U}
U^{-1})$ for this configuration. Since the integral is invariant
under continuous deformations, we can take very small values of
$\rho$ for this calculation. In this case, (\ref{6}) gives
\beqar 
U&& =1, ~~~~~~~~~~~~~~~~~~~~~~~~~~~n>0\nonumber\\
&&=\left( \matrix {e^{i\alpha} &0\cr
                   0&1\cr}\right),~~~~~~~~~~~~~~~n=0 \label{9}
\eeqar 
where 
\beq
e^{i\alpha (t)} = \frac{t-i\rho}{t+i\rho}.
\eeq
Clearly, $\Delta \alpha = \alpha (\infty )-
\alpha (-\infty ) =2\pi$ in this case, giving $\int dt~\Tr
({\dot U}U^{-1}) = 2\pi i$. We have a configuration of winding
number 1 of $\Pi_1 (U(N))$ for $N \gg \rho^2/\theta$.

As $\theta$ goes to zero, the transformation (\ref{6}) becomes a
winding number 1 element of
$\Pi_3 (SU(2))$. In fact, by using coherent states, we find for
$\theta \rightarrow 0$ 
\beq
U_{ab} \rightarrow
\left( {x^2-\rho^2 \over x^2
+\rho^2 } -{2i\rho {\vec x} \cdot {\vec \sigma}\over x^2
+\rho^2}\right)_{ab} \label{10}
\eeq
where $x^2 =2z\bz +t^2$, $x^3 =t$ and $\sigma_i$ are the
Pauli matrices. As
$\theta$ goes to zero, we see that $U$ goes over to the smooth
configuration of winding number 1 of $\Pi_3 (SU(2))$
corresponding to the stereographic map of the three-sphere. 

We can also take the $\theta \rightarrow 0$ limit considering
$U$ to be  a $U(1)$-type transformation. Embedding the states
in a single Fock space as in (\ref{8}), we have
\beqar 
U(t) ~\vert 2n\ra &&= A_n \vert 2n \ra ~+~ B_n \vert
2n-1\ra \nonumber\\ 
U(t) ~\vert 2n+1\ra && = A^*_{n+1} \vert
2n+1\ra ~+~ B_n \vert 2n+2\ra \label{11}
\eeqar 
Since $\vert n-1\ra \sim e^{i\varphi} \vert n\ra$,
$x^1 +i x^2 \sim z = r~ e^{i\varphi}$, we see that $U(t)$
has a part that goes like $e^{i\varphi}$ or $e^{-i\varphi}$ for
even and odd states respectively. Similarly, it has a part that
goes as ${(z\bz - (\rho \pm it )^2 )/( z\bz +\rho^2+t^2) }$
for even and odd states. Thus, considered as a single
$U(1)$-type transformation, it has a highly oscillatory
behaviour on scales $\Delta r^2 \sim \theta$, and becomes
singular as $\theta \rightarrow 0$.  Strictly in the commutative
limit, therefore, we do not need to require invariance under
such transformations and there is no reason for quantization of
the level number. However, if the commutative Abelian theory is
viewed as the small $\theta$-limit of the noncommutative theory,
quantization persists.

\vskip .2in
\leftline{\bf Acknowledgments}

We would like to thank Bogdan Morariu for interesting discussions.
This work was supported in part by the National Science
Foundation grant PHY-0070883 and a CUNY
Collaborative Incentive Research Grant.

\end{document}